# Title

Combining n-MOS Charge Sensing with p-MOS Silicon Hole Double Quantum Dots in a CMOS platform

# Authors

Ik Kyeong Jin,[1] Krittika Kumar,[1] Matthew J. Rendell,[1] Jonathan Y. Huang,[2,3] Chris C. Escott,[2,3] Fay E. Hudson,[2,3] Wee Han Lim,[2,3] Andrew S. Dzurak,[2,3] Alexander R. Hamilton,[1] and Scott D. Liles[1]*

# Affiliations

[1]School of Physics, The University of New South Wales, Sydney NSW 2052, Australia.
[2]School of Electrical Engineering and Telecommunications, The University of New South Wales, Sydney NSW 2052, Australia.
[3]Diraq, Sydney, NSW 2052, Australia.

* Corresponding author: s.liles@unsw.edu.au

# Abstract

Holes in silicon quantum dots are receiving significant attention due to their potential as fast, tunable, and scalable qubits in semiconductor quantum circuits. Despite this, challenges remain in this material system including difficulties using charge sensing to determine the number of holes in a quantum dot, and in controlling the coupling between adjacent quantum dots. In this work, we address these problems by fabricating an ambipolar complementary metal-oxide-semiconductor (CMOS) device using multilayer palladium gates. The device consists of an electron charge sensor adjacent to a hole double quantum dot. We demonstrate control of the spin state via electric dipole spin resonance (EDSR). We achieve smooth control of the inter-dot coupling rate over two orders of magnitude and use the charge sensor to perform spin-to-charge conversion to measure the hole singlet-triplet relaxation time of 11 μs for a known hole occupation. These results provide a path towards improving the quality and controllability of hole spin-qubits.

# Introduction

The unique properties of hole spins make them a promising candidate for implementing scalable semiconductor-based qubits. The strong spin-orbit interaction (SOI) for holes provides a coupling mechanism between hole-spins and electric fields (*1, 2*) allowing hole-spins to be rapidly manipulated by simply applying an AC electric field to a local electrode (*3*). Recent experiments using holes in quantum dots have shown fast electric spin manipulation ($f_{Rabi}$ > 100 MHz (*4–6*)), demonstrated single spin-qubit devices in silicon (*7*) and four-qubit devices in germanium (*8*). Furthermore, the SOI of holes in quantum dots is highly tunable in-situ through parameters including the confinement shape, magnetic-field orientation or spatial position with respect to non-uniform strain (*9, 10*). Therefore, recent calculations, confirmed by experiments, suggested it is possible to engineer "sweet spots" for holes, where rapid spin control is possible, while simultaneously minimizing the effect of charge noise induced decoherence (*11–13*).

In silicon, the fabrication and processing of p-type quantum devices is far less developed compared with the more conventional n-type quantum devices. Materials challenges such as the high contact resistance, charge noise (*14*) and low mobility of p-type devices result in much lower bandwidth and worse signal-to-noise ratio for p-type quantum devices, causing significant limitations for p-type qubit readout (see also Supplementary Material I). The main strategies to address hole qubit readout have been to develop sensing methods that avoid the known drawbacks of p-type devices (such as gate-based dispersive readout (*15–17*)) or to continue to develop and optimize p-type processing (*18–20*). However, there is a potential shortcut: take advantage of the ambipolar capabilities of CMOS silicon to implement a p-type qubit with a high-quality n-type charge sensor.

CMOS devices commonly combine both n-type and p-type capabilities within the same device. However, in quantum devices based on CMOS technology, devices are typically unipolar. The limited studies of ambipolar quantum dots typically used transport measurements of a single quantum dot connected to both n-type and p-type ohmics (*14, 21–25*). Recently, Souza de-Almeda et al. (*26*) demonstrated a device with an n-type dot to read out the charge state of a p-type dot (and vice versa). Now the open question is the feasibility of using ambipolar charge sensing in spin qubit devices at the single hole level.

In this work we demonstrate an improved hole spin qubit architecture by (a) combining n-type charge sensor with a p-type double dot and (b) using Pd to fabricate multilayer quantum dot devices instead of using Al. The presented device architecture enables readout of hole spin relaxation in a silicon hole MOS platform via an adjacent charge sensor. The device shows a hole singlet-triplet relaxation time of 11µs, which is reported for a known double dot hole occupation of (2,8). Finally, the device demonstrates the capability for electrically driven spin resonance, while showing smooth controllability of the inter-dot coupling rate over two orders of magnitude.

## Results

### Combining n-MOS charge sensing with p-MOS quantum dots

Figure 1A shows a scanning electron microscope (SEM) image of the CMOS device design. Using palladium allowed finer gate features than previous aluminium gates (*18, 27*) which will be important for inter-dot coupling control discussed later (See Methods for detail). Full characterisation details of the mobility and threshold voltages for the n-MOS and p-MOS are given in Supplementary Material I. The left side of the device has p-type ohmic contacts to form a hole double quantum dot (blue), while the right side of the device has n-type ohmic contacts to form a single-electron dot charge sensor (red). This geometry allows the formation of an electron dot close to a hole double quantum dot (Fig. 1B).

We first demonstrate simultaneous operation of the hole double quantum dot and the adjacent single-electron transistor (SET) charge sensor. Fig. 1C shows the measured dc current through the p-type channel. In the bottom left region of the stability diagram several bias triangles are observed, consistent with the formation of a hole double quantum dot. As the biases on P1 and P2 are made more positive the bias triangles lose visibility since voltages applied on P1 and P2 gates control the hole tunnelling rate, resulting in a precipitous drop in the current.

Figure 1D, on the other hand, shows the current through the adjacent SET ($I_{CS}$), measured simultaneously to Fig. 1C. The charge transition lines show the typical honeycomb pattern expected for a double quantum dot controlled by gates P1 and P2. The grey rectangular boxes in Fig. 1C and 1D indicate the bias triangles and corresponding triple points which were observed simultaneously in transport and charge sensing. Unlike the p-type transport of Fig. 1C, the charge sensor is able to observe the charge transitions of the double quantum dot all the way down to the last hole. This allows confirmation of the exact hole occupation of the double dot, which is necessary to understand its orbital structure and hence its spin-orbit interaction.

Figure 1E shows measurements of the charge sensor when the double quantum dot is operating in the last hole regime. Due to the asymmetry of the SET location with respect to the P1 and P2 dot, the sensor is able to discriminate between the (2,0) and (1,1) states due to the charge dipole (also see Supplementary Material II). This is useful for enhancing the signal when performing Pauli Spin Blockade (PSB) based spin-to-charge conversion discussed later.

### Electric control of the hole spin state

Next, we demonstrate electric control of the hole spin state using measurements of the current through the double dot. The device was tuned to a bias triangle where the base current was suppressed by spin blockade (Fig. 2A). The PSB can be lifted by applying a microwave tone to plunger gate P1, where the frequency of the tone is equal to the Zeeman splitting (Fig. 2B). Increasing the magnetic field increases the microwave frequency needed to match the Zeeman splitting (Fig. 2C). These results demonstrate that it is possible to implement ambipolar charge sensing using a spin-qubit device layout, without limiting the ability to detect and perform spin manipulation.

### Smooth control of inter-dot tunnel coupling rate

A main motivation for using electrically defined quantum dots as qubits is the promise of smooth in-situ control of the inter-dot tunnel coupling ($t_c$). However, in silicon MOS quantum dots, reliable in-situ control of $t_c$ has remained a challenge and has only recently been achieved for n-type silicon devices (*28*). We now use the n-type charge sensor to demonstrate smooth control of $t_c$ for a p-type planar hole double quantum dot. The voltage applied to the J-gate ($J_g$) allows control of inter-dot coupling. Three representative stability diagrams are shown in Fig. 3: (A) strong inter-dot coupling (electrons occupying molecular orbitals that span both dots) (B) intermediate inter-dot coupling and (C) weak inter-dot coupling (localized double dot). Fig. 3D shows the measured $t_c$ as a function of $J_g$, measured at the true (1,1) to (0,2) transition. The interdot tunnel coupling rates are calculated from the measured charge distribution as a function of detuning between the P1 and P2 dot potentials (*29*). Details can be found in Supplementary Material III. The data show smooth control of interdot tunnel coupling rates over two orders of magnitude.

### Measurement of singlet-triplet relaxation time

To verify that the charge sensor is capable of spin-to-charge conversion, we measured the singlet-triplet relaxation time ($T_{ST}$) of the double-hole quantum dot using the n-type charge sensor. Firstly, we located a charge transition that shows PSB by observing transport measurements at 3mV of source-drain bias (see Supplementary Material IV). The (2,8) - (1,9) transition (which we analyse as an effective "(2,0)" to "(1,1)" transition, assuming standard two-charge singlet-triplet terminology) showed clear spin blockade with a singlet-triplet energy splitting of 460 µeV (see supplementary Material IV). $T_{ST}$ measurements were performed using the (2,8) - (1,9) transition rather than the true (2,0) - (1,1) transition since long reservoir-to-dot tunnel rates occur at the last hole ($T_{load} \gg T_{ST}$), due to effect of strongly depleting plunger gates (see Fig 1C).

To detect the spin blockade using the charge sensor a three-stage gate pulse sequence is employed (*30*). The cyclical three-stage pulse sequence consisted of Empty (E) - Load (L) - Measure (M) stages as shown in the schematics in Fig. 4D. The Empty stage resets the double dot into (1,8), and the Load stage loads randomly loads a (1,9) singlet or triplet configuration. When the M stage lies within the spin blockaded region of the (2,8) charge configuration, the triplet (1,9) states are blocked, while the (1,9) singlet states quickly tunnel into (2,8), allowing spin-to-charge conversion. To enhance the PSB signal we alternate between clockwise E-L-M and counter-clockwise L-E-M sequences to allow a differential lock-in measurement (the resultant is denoted as $I_{CS,AC}$ in Fig. 4B and C). The PSB region in Fig. 4B confirms that we can use the electron charge sensor to detect the hole spin state (see also Supplementary Material V).

Next, we measured the singlet-triplet relaxation time within the PSB region (marked by an orange star on Fig. 4B). This measurement was performed by increasing the pulse period, while keeping the ratio of E:L:M stages fixed at 1:1:6, so that the triplet states have more time to relax into the singlet states. Fig. 4E shows the sensor signal as the pulse period is increased for three different in-plane magnetic field values. We fit the decay in the sensor signal to extract $T_{ST}$ (see Supplementary Material VI) and obtained a maximum $T_{ST}$ of 11 ± 3µs at zero magnetic field (Fig. 4D). The singlet-triplet relaxation time dramatically falls as a function of in-plane magnetic field. This is consistent with the reduced visibility of the spin-blockade signal in Fig. 4C, as well as the recent results for holes in planar Ge quantum dots (*31*).

## Discussion

We demonstrated integration of an n-type SET as a charge sensor with a p-type double quantum dot in a planar MOS structure. The charge state of the p-type double quantum dot can be simultaneously measured by p-type transport as well as n-type charge sensing. The charge sensor is able to sense down to the last few holes in the double quantum dot. The CMOS charge sensor is able to differentiate states with the same total charge occupation, e.g. (2,0) state and (1,1) state. Singlet-triplet relaxation time was measured at various magnetic fields using the CMOS charge sensor. We demonstrated smooth control of the interdot barrier tunnel coupling rates over two orders of magnitude as well as electrical control of the spin state of the hole double dot via EDSR.

Low bandwidth of the sensor (~100 ms integration time), and the precipitous drop in $T_{ST}$ due to increasing magnetic field prevented ability to investigate spin physics at time scales faster than 1 μs including the single shot measurement. Future work will improve the measurement bandwidth with RF reflectometry using the low resistance n-type sensor (*32*) combined with latched readout (*33*) (preliminary results are shown in Supplementary VII). This work provides a step towards realizing high speed single hole qubits with high performance charge sensors.

## Materials and Methods

### Device fabrication
The device is fabricated on a high resistivity natural silicon wafer with a high quality 5.9 nm $SiO_2$ gate oxide. P+ (n+) ohmic regions were prepared by boron (phosphorus) diffusion. The planar gate structure was fabricated using a multilayer Pd metal and an atomic layer deposition (ALD) 2 nm $Al_2O_3$ gate stack (*27*).

### Experimental setup
Plunger gates P1 and P2 were negatively biased to form a double hole quantum dot and ST was positively biased to form a single-electron quantum dot. The measured lever arms and calculated capacitive matrix for this device are discussed in more detail in the Supplementary material VIII. The currents in the p-type and n-type channels are simultaneously measured using low noise current-to-voltage converters and standard low frequency dual lock-in techniques with applied source-drain bias of 3 mV (*34*). All experiments were performed at cryogenic temperatures using a BlueFors dilution fridge with a hole/electron temperature of ≈ 100 mK.

# Figures and Tables

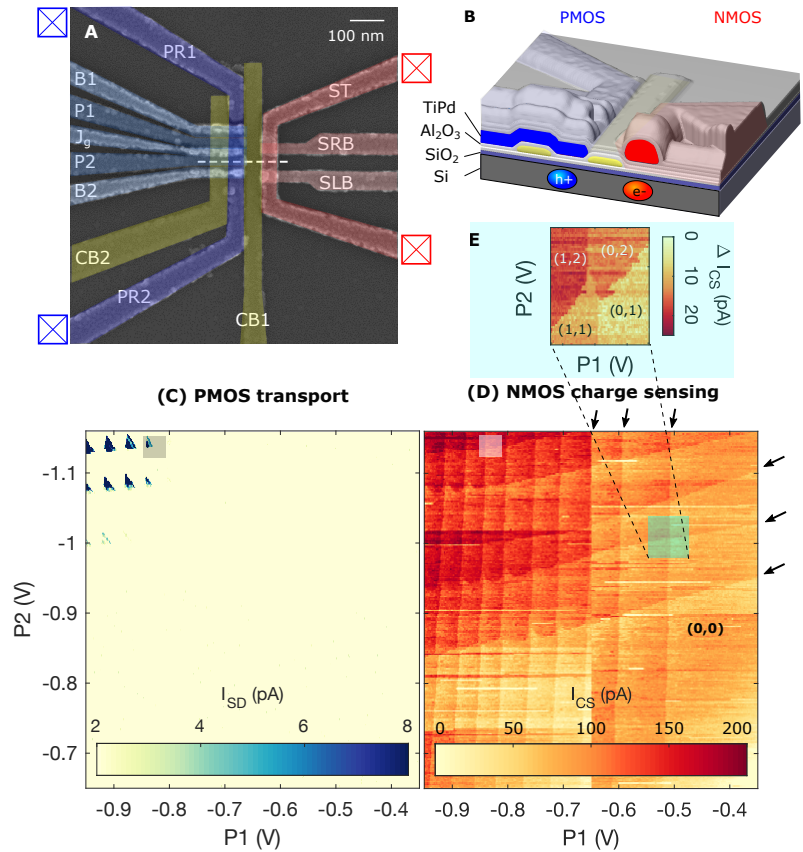

**Fig. 1. CMOS ambipolar charge sensing device.** (**A**) A false-colored scanning electron microscopy (SEM) image of an ambipolar device with planar MOS structure. The left half of the device is able to form a double hole quantum dot with p-type ohmic contacts, whereas the right half of the device is able to form a single-electron quantum dot with n-type ohmic contacts. (**B**) Cross section of the 3D-model generated ambipolar device. The fabricated device is able to form a hole quantum dot next to an electron quantum dot close to each other, which enables mutual charge sensing of n-type and p-type devices. (**C**) Bias triangles of a double hole quantum dot measured by p-type transport with $V_{SD}$ = 3 mV and (**D**) simultaneously measured stability diagram with an n-type charge sensor. Grey rectangles highlight the same charge transition region on the two maps. Charge sensing allows detection of the charge occupation down to the last hole, which is not possible in transport as the barriers become too opaque. The last three transitions are marked by black arrows. (**E**) Zoomed in stability diagram at the (1,1) – (0,2) transition showing four distinct current levels. The vertical line at P1 = -0.65 V in Fig. 1D is an artifact due to re-tuning the sensor over this large map.

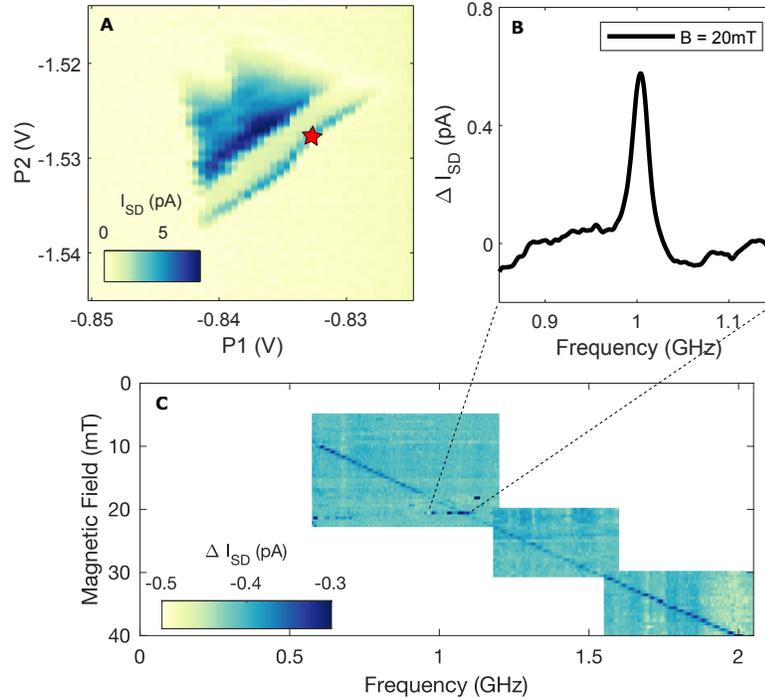

**Fig. 2. Electric dipole spin resonance (EDSR) of the hole double quantum dot.** (**A**) A bias triangle of the double hole quantum dot. Pauli spin blockade (PSB) is observed at the base of the triangle (red star). (**B**) Microwave is applied to a hole dot to lift PSB. The PSB is lifted only when the applied microwave energy matches the Zeeman splitting of the system. (**C**) The EDSR response which demonstrates rapid electric control of double hole quantum dot spin states. Measured g-factor from this experiment is 1.12.

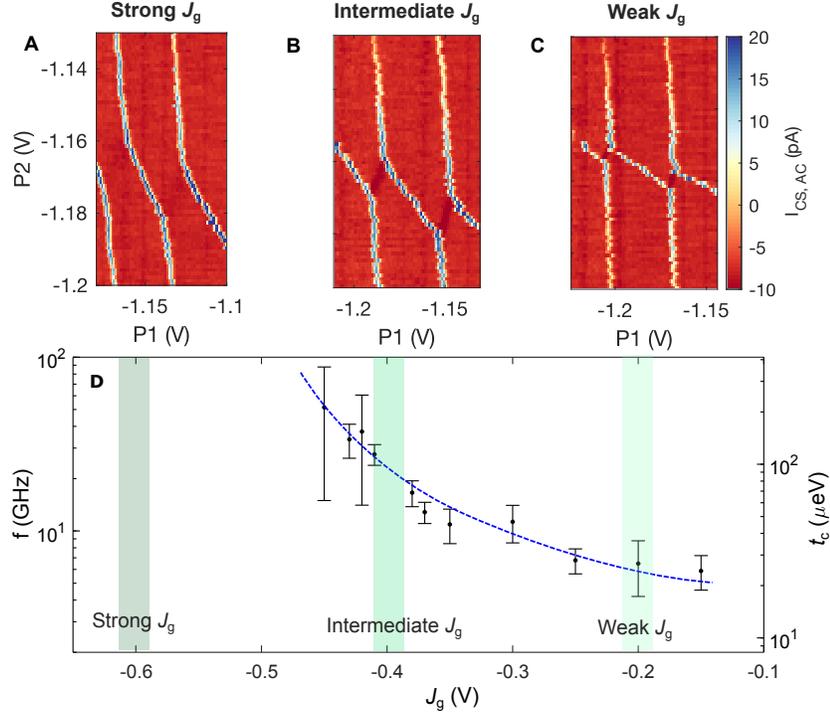

**Fig. 3. Smooth control of interdot tunnel coupling rate of the hole double quantum dot.** (**A-C**) Stability diagrams of the hole double quantum dot at different interdot barrier voltages $J_g$ measured by the n-type charge sensor. Change in slope of the charge transition lines and broadening of the interdot coupling line shows that interdot coupling rates are different at each regime. (**D**) Measured interdot coupling rate $t_c$ as a function of $J_g$ at the true (1,1) to (0,2) transition. The blue line is guide to the eye. At strong J g regime, interdot tunnelling rates are too fast to measure ($\gg$ 100 GHz). Interdot tunnelling rates are calculated using the measured width of the detuning peak ((*29*)).

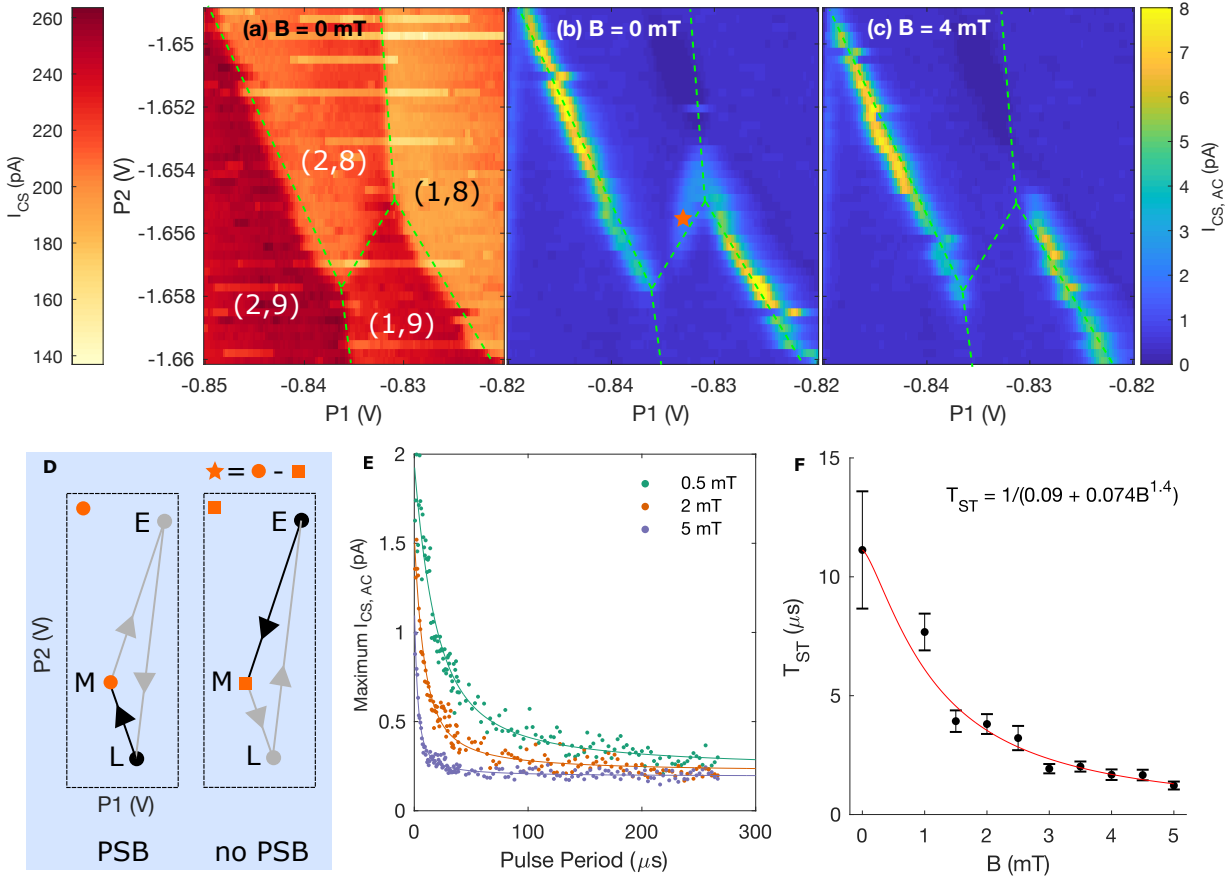

**Fig. 4. Singlet-triplet relaxation time ($T_{ST}$) measurement of a silicon hole using the adjacent electron charge sensor.** (**A**) Stability diagram at the zero magnetic field. CB-gate voltages are set at to 0.5 V, compared to 0.14 V in Fig 1. (**B, C**) A finite lock-in signal is measured near zero detuning on the "(2,0)" region of the stability diagram. When the magnetic field is increased this signal disappears due to increased singlet-triplet relaxation. (**D**) A cyclical three stage pulse E-L-M is used to observe Pauli Spin blocked region on the "(2,0)" side of the interdot transition. The DC voltage of P1 and P2 is scanned around the (2,8) to (1,9) (i.e., "(2,0)" to "(1,1)") transition while keeping the pulse sequence fixed. The timing of each stage E: L: M was 0.5 µs: 0.5 µs: 3 µs causing the signal to be dominated by the charge sensor current at the M point. We note that the lock-in signal also picks up the charge transitions, which are independent of magnetic field amplitude (*35*). (**E**) AC current at the PSB region (orange star on Fig. 4B) at different pulse periods and magnetic field. (**F**) Singlet-triplet relaxation time is extracted at different magnetic fields. Fitting function is displayed as a red line.

## Acknowledgments

**Funding:** This work was funded by the Australian Research Council (Grants No. DP150100237, No. DP200100147, and No. FL190100167) and the U.S. Army Research Office (Grant No. W911NF-17-1-0198). Devices were made at the New South Wales node of the Australian National Fabrication Facility.

**Author contributions:** ARH, ASD, and SDL conceptualized the project. FEH and WHL fabricated the devices. SDL and IKJ performed the experiments. SDL, IKJ, KK and MJR analyzed the data. IKJ, SDL, KK, MJR, FEH, WHL, ASD and ARH contributed to discussions on the experimental results. JYH, CCE performed electrostatic simulation of the devices. IKJ, JYH, KK, CCE and SDL visualized the data. IKJ, SDL, and ARH wrote the manuscript with reviews from all authors. SDL, ASD, and ARH supervised the project.

**Competing interests:** ASD is CEO and a director of Diraq Pty Ltd. All other authors declare no competing interests.

**Data and materials availability:** All data needed to evaluate the conclusions in the paper are present in the paper and/or the Supplementary Materials. Additional data related to this paper may be requested from the authors.

# References

1. E. I. Rashba, Al. L. Efros, Orbital Mechanisms of Electron-Spin Manipulation by an Electric Field. *Phys. Rev. Lett.* **91**, 126405 (2003).

2. M. Duckheim, D. Loss, Electric-dipole-induced spin resonance in disordered semiconductors. *Nat. Phys.* **2**, 195–199 (2006).

3. S. Nadj-Perge, S. M. Frolov, E. P. A. M. Bakkers, L. P. Kouwenhoven, Spin–orbit qubit in a semiconductor nanowire. *Nature*. **468**, 1084–1087 (2010).

4. L. C. Camenzind, S. Geyer, A. Fuhrer, R. J. Warburton, D. M. Zumbühl, A. V. Kuhlmann, A hole spin qubit in a fin field-effect transistor above 4 kelvin. *Nat. Electron.* **5**, 178–183 (2022).

5. H. Watzinger, J. Kukučka, L. Vukušić, F. Gao, T. Wang, F. Schäffler, J.-J. Zhang, G. Katsaros, A germanium hole spin qubit. *Nat. Commun.* **9**, 3902 (2018).

6. N. W. Hendrickx, D. P. Franke, A. Sammak, G. Scappucci, M. Veldhorst, Fast two-qubit logic with holes in germanium. *Nature*. **577**, 487–491 (2020).

7. R. Maurand, X. Jehl, D. Kotekar-Patil, A. Corna, H. Bohuslavskyi, R. Laviéville, L. Hutin, S. Barraud, M. Vinet, M. Sanquer, S. De Franceschi, A CMOS silicon spin qubit. *Nat. Commun.* **7** (2016), doi:10.1038/ncomms13575.

8. N. W. Hendrickx, W. I. L. Lawrie, M. Russ, F. van Riggelen, S. L. de Snoo, R. N. Schouten, A. Sammak, G. Scappucci, M. Veldhorst, A four-qubit germanium quantum processor. *Nature*. **591**, 580–585 (2021).

9. S. D. Liles, F. Martins, D. S. Miserev, A. A. Kiselev, I. D. Thorvaldson, M. J. Rendell, I. K. Jin, F. E. Hudson, M. Veldhorst, K. M. Itoh, O. P. Sushkov, T. D. Ladd, A. S. Dzurak, A. R. Hamilton, Electrical control of the g tensor of the first hole in a silicon MOS quantum dot. *Phys. Rev. B*. **104**, 235303 (2021).

10. F. N. M. Froning, L. C. Camenzind, O. A. H. van der Molen, A. Li, E. P. A. M. Bakkers, D. M. Zumbühl, F. R. Braakman, Ultrafast hole spin qubit with gate-tunable spin–orbit switch functionality. *Nat. Nanotechnol.* **16**, 308–312 (2021).

11. Z. Wang, E. Marcellina, Alex. R. Hamilton, J. H. Cullen, S. Rogge, J. Salfi, D. Culcer, Optimal operation points for ultrafast, highly coherent Ge hole spin-orbit qubits. *Npj Quantum Inf.* **7**, 54 (2021).

12. S. Bosco, B. Hetényi, D. Loss, Hole Spin Qubits in Si FinFETs With Fully Tunable Spin-Orbit Coupling and Sweet Spots for Charge Noise. *PRX Quantum*. **2**, 010348 (2021).

13. N. Piot, B. Brun, V. Schmitt, S. Zihlmann, V. P. Michal, A. Apra, J. C. Abadillo-Uriel, X. Jehl, B. Bertrand, H. Niebojewski, L. Hutin, M. Vinet, M. Urdampilleta, T. Meunier, Y.-M. Niquet, R. Maurand, S. D. Franceschi, A single hole spin with enhanced coherence in natural silicon. *Nat. Nanotechnol.* (2022), doi:10.1038/s41565-022-01196-z.

14. A. C. Betz, M. F. Gonzalez-Zalba, G. Podd, A. J. Ferguson, Ambipolar quantum dots in intrinsic silicon. *Appl. Phys. Lett.* **105**, 153113 (2014).

15. J. I. Colless, A. C. Mahoney, J. M. Hornibrook, A. C. Doherty, H. Lu, A. C. Gossard, D. J. Reilly, Dispersive Readout of a Few-Electron Double Quantum Dot with Fast rf Gate Sensors. *Phys. Rev. Lett.* **110** (2013), doi:10.1103/PhysRevLett.110.046805.

16. A. Crippa, R. Ezzouch, A. Aprá, A. Amisse, R. Laviéville, L. Hutin, B. Bertrand, M. Vinet, M. Urdampilleta, T. Meunier, M. Sanquer, X. Jehl, R. Maurand, S. De Franceschi, Gate-reflectometry dispersive readout and coherent control of a spin qubit in silicon. *Nat. Commun.* **10**, 2776 (2019).

17. R. Ezzouch, S. Zihlmann, V. P. Michal, J. Li, A. Aprá, B. Bertrand, L. Hutin, M. Vinet, M. Urdampilleta, T. Meunier, X. Jehl, Y.-M. Niquet, M. Sanquer, S. De Franceschi, R. Maurand, Dispersively Probed Microwave Spectroscopy of a Silicon Hole Double Quantum Dot. *Phys. Rev. Appl.* **16**, 034031 (2021).

18. S. D. Liles, R. Li, C. H. Yang, F. E. Hudson, M. Veldhorst, A. S. Dzurak, A. R. Hamilton, Spin and orbital structure of the first six holes in a silicon metal-oxide-semiconductor quantum dot. *Nat. Commun.* **9** (2018), doi:10.1038/s41467-018-05700-9.

19. R. Li, N. I. D. Stuyck, S. Kubicek, J. Jussot, B. T. Chan, F. A. Mohiyaddin, A. Elsayed, M. Shehata, G. Simion, C. Godfrin, Y. Canvel, Ts. Ivanov, L. Goux, B. Govoreanu, I. P. Radu, "A flexible 300 mm integrated Si MOS platform for electron- and hole-spin qubits exploration" in *2020 IEEE International Electron Devices Meeting (IEDM)* (2020), p. 38.3.1-38.3.4.

20. S. Geyer, L. C. Camenzind, L. Czornomaz, V. Deshpande, A. Fuhrer, R. J. Warburton, D. M. Zumbühl, A. V. Kuhlmann, Self-aligned gates for scalable silicon quantum computing. *Appl. Phys. Lett.* **118**, 104004 (2021).

21. R. Martel, V. Derycke, C. Lavoie, J. Appenzeller, K. K. Chan, J. Tersoff, Ph. Avouris, Ambipolar Electrical Transport in Semiconducting Single-Wall Carbon Nanotubes. *Phys. Rev. Lett.* **87**, 256805 (2001).

22. J. Güttinger, C. Stampfer, F. Libisch, T. Frey, J. Burgdörfer, T. Ihn, K. Ensslin, Electron-Hole Crossover in Graphene Quantum Dots. *Phys. Rev. Lett.* **103**, 046810 (2009).

23. J. C. H. Chen, O. Klochan, A. P. Micolich, K. Das Gupta, F. Sfigakis, D. A. Ritchie, K. Trunov, D. Reuter, A. D. Wieck, A. R. Hamilton, Fabrication and characterisation of gallium arsenide ambipolar quantum point contacts. *Appl. Phys. Lett.* **106**, 183504 (2015).

24. A. V. Kuhlmann, V. Deshpande, L. C. Camenzind, D. M. Zumbühl, A. Fuhrer, Ambipolar quantum dots in undoped silicon fin field-effect transistors. *Appl. Phys. Lett.* **113**, 122107 (2018).

25. J. Duan, J. S. Lehtinen, M. A. Fogarty, S. Schaal, M. Lam, A. Ronzani, A. Shchepetov, P. Koppinen, M. Prunnila, F. Gonzalez-Zalba, J. J. L. Morton, Dispersive readout of reconfigurable ambipolar quantum dots in a silicon-on-insulator nanowire. *Appl. Phys. Lett.* **118**, 164002 (2021).

26. A. J. Sousa de Almeida, A. M. Seco, T. van den Berg, B. van de Ven, F. Bruijnes, S. V. Amitonov, F. A. Zwanenburg, Ambipolar charge sensing of few-charge quantum dots. *Phys. Rev. B*. **101**, 201301 (2020).

27. R. Zhao, T. Tanttu, K. Y. Tan, B. Hensen, K. W. Chan, J. C. C. Hwang, R. C. C. Leon, C. H. Yang, W. Gilbert, F. E. Hudson, K. M. Itoh, A. A. Kiselev, T. D. Ladd, A. Morello, A. Laucht, A. S. Dzurak, Single-spin qubits in isotopically enriched silicon at low magnetic field. *Nat. Commun.* **10**, 5500 (2019).

28. H. G. J. Eenink, L. Petit, W. I. L. Lawrie, J. S. Clarke, L. M. K. Vandersypen, M. Veldhorst, *Nano Lett.*, in press, doi:10.1021/acs.nanolett.9b03254.

29. L. DiCarlo, H. J. Lynch, A. C. Johnson, L. I. Childress, K. Crockett, C. M. Marcus, M. P. Hanson, A. C. Gossard, Differential Charge Sensing and Charge Delocalization in a Tunable Double Quantum Dot. *Phys. Rev. Lett.* **92**, 226801 (2004).

30. J. R. Petta, A. C. Johnson, A. Yacoby, C. M. Marcus, M. P. Hanson, A. C. Gossard, Pulsed-gate measurements of the singlet-triplet relaxation time in a two-electron double quantum dot. *Phys. Rev. B*. **72**, 161301 (2005).

31. D. Jirovec, A. Hofmann, A. Ballabio, P. M. Mutter, G. Tavani, M. Botifoll, A. Crippa, J. Kukucka, O. Sagi, F. Martins, J. Saez-Mollejo, I. Prieto, M. Borovkov, J. Arbiol, D. Chrastina, G. Isella, G. Katsaros, A singlet-triplet hole spin qubit in planar Ge. *Nat. Mater.* **20**, 1106–1112 (2021).

32. R. J. Schoelkopf, The Radio-Frequency Single-Electron Transistor (RF-SET): A Fast and Ultrasensitive Electrometer. *Science*. **280**, 1238–1242 (1998).

33. P. Harvey-Collard, B. D'Anjou, M. Rudolph, N. T. Jacobson, J. Dominguez, G. A. Ten Eyck, J. R. Wendt, T. Pluym, M. P. Lilly, W. A. Coish, M. Pioro-Ladrière, M. S. Carroll, High-Fidelity Single-Shot Readout for a Spin Qubit via an Enhanced Latching Mechanism. *Phys. Rev. X*. **8**, 021046 (2018).

34. J. M. Elzerman, R. Hanson, L. H. Willems van Beveren, L. M. K. Vandersypen, L. P. Kouwenhoven, Excited-state spectroscopy on a nearly closed quantum dot via charge detection. *Appl. Phys. Lett.* **84**, 4617–4619 (2004).

35. A. P. Higginbotham, T. W. Larsen, J. Yao, H. Yan, C. M. Lieber, C. M. Marcus, F. Kuemmeth, Hole Spin Coherence in a Ge/Si Heterostructure Nanowire. *Nano Lett.* **14**, 3582–3586 (2014).

36. C. H. Yang, W. H. Lim, F. A. Zwanenburg, A. S. Dzurak, Dynamically controlled charge sensing of a few-electron silicon quantum dot. *AIP Adv.* **1**, 042111 (2011).

37. J. R. Petta, A. C. Johnson, J. M. Taylor, E. A. Laird, A. Yacoby, M. D. Lukin, C. M. Marcus, M. P. Hanson, A. C. Gossard, Coherent Manipulation of Coupled Electron Spins in Semiconductor Quantum Dots. *Science* (2005).

# Supplementary Materials for

## Combining n-MOS Charge Sensing with p-MOS Silicon Hole Double Quantum Dots in a CMOS platform

Ik Kyeong Jin, Krittika Kumar, Matthew J. Rendell, Jonathan Y. Huang, Chris C. Escott, Fay E. Hudson, Wee Han Lim, Andrew S. Dzurak, Alexander R. Hamilton, and Scott D. Liles*

*Corresponding author. Email: s.liles@unsw.edu.au

# I. MOBILITY MEASUREMENTS ON TEST MOSFET DEVICES

We fabricated two terminal MOSFETs with the same gate stacks as the fabricated CMOS charge sensing quantum dot device (Figure S1a). We compared threshold voltage and mobility of n-type devices compared to p-type devices on the same chip at four Kelvin, with different gate materials (Al, Ti) and different oxides ($Al_2O_3$, $HfO_2$). Figure S1b-e shows that the n-type MOSFETs have higher mobility compared to the p-type MOSFETs, which demonstrates an advantage of using an n-type device as a charge sensor in terms of mobility.

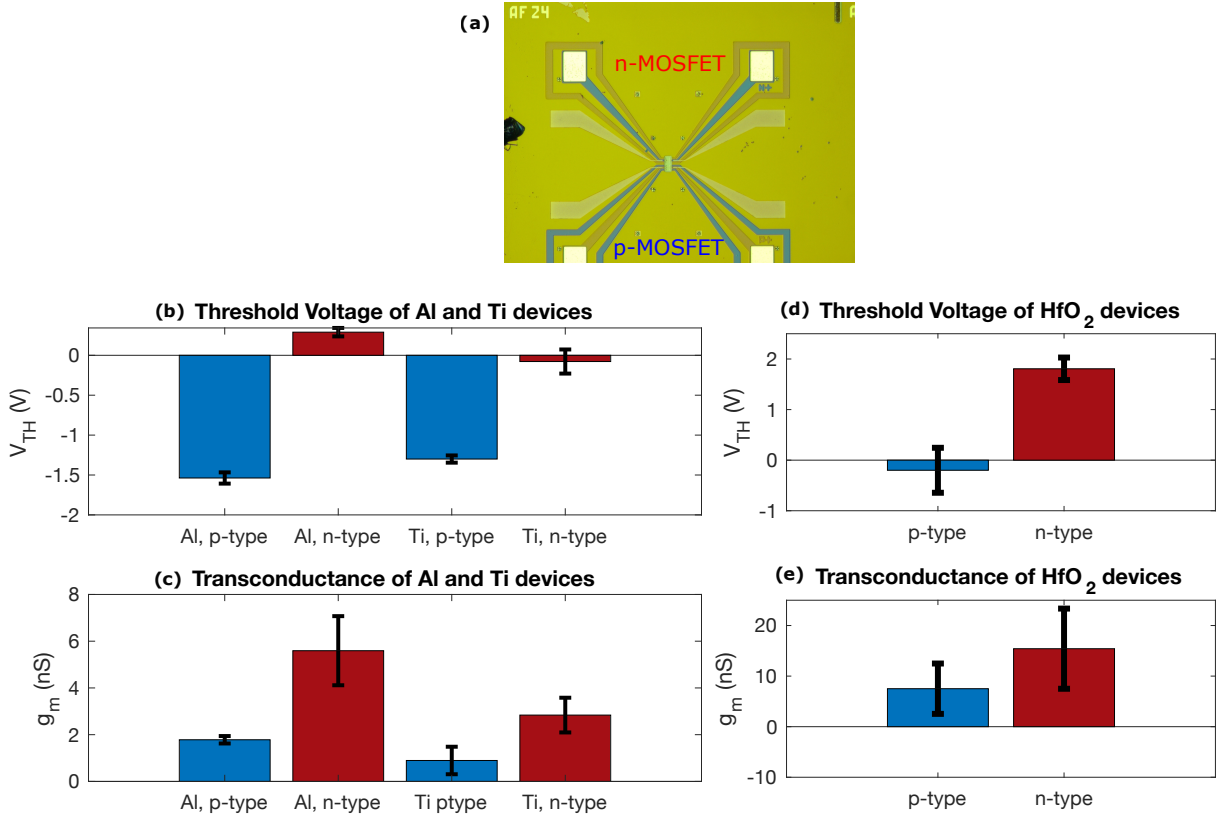

**FIG. S1. Mobility and threshold voltage measurements on test MOSFET devices.** (a) Optical image of a MOSFET device to characterize electron and hole mobility on the same chip. (b-e) Measured threshold voltage ($V_{TH}$) and transconductance ($g_m$) of the test MOSFET devices. Transconductance, which is proportional to the mobility of the n-type devices, is higher than p-type devices in all the batches we fabricated.

## II. HISTOGRAM OF THE CHARGE SENSOR CURRENT AT THE LAST HOLE REGIME

We measured the charge sensor current around the true (2,0) - (1,1) transition. The change of charge sensing current $\Delta I_{CS}$ is obtained after compensating the capacitive coupling of the plunger gates P1 and P2 to the sensor dot (*36*). Four separate peaks show that the electron charge sensor is sensitive enough to distinguish between the four distinct charge occupations of the nearby hole quantum dot. In particular we demonstrate a ∼10 pA charge sensor signal between the (1,1) and (2,0) configurations, which is key for spin-to-charge conversion.

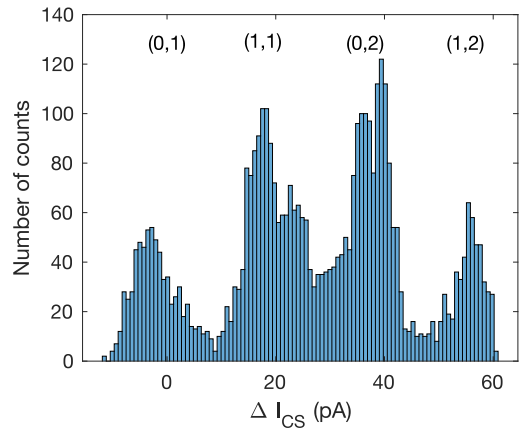

**FIG. S2. Charge sensor current measured at the last hole regime.** Four distinct current levels demonstrate ability of the fabricated charge sensor to measure charge occupation of the adjacent hole double quantum dot.

## III. INTERDOT TUNNEL COUPLING RATE CALCULATION

The interdot tunnel coupling rate is measured from the width of the detuning between the P1 and P2 dot potentials (Figure S3). The equation used for the fitting is Ref. (*29*) equation 2.

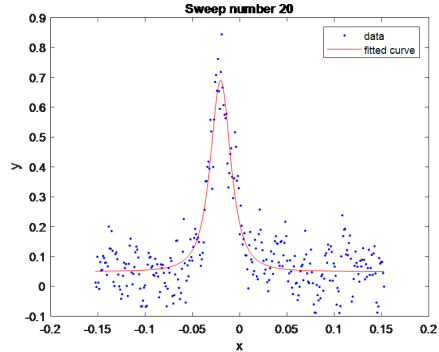

**FIG. S3. Normalized data of a detuning peak.** The lever arms are calculated from the width of the detuning peak.

Firstly we performed temperature dependent fits to get lever arms of the fitting function. The quadrature of in phase and out of phase signal was used as the amplitude.

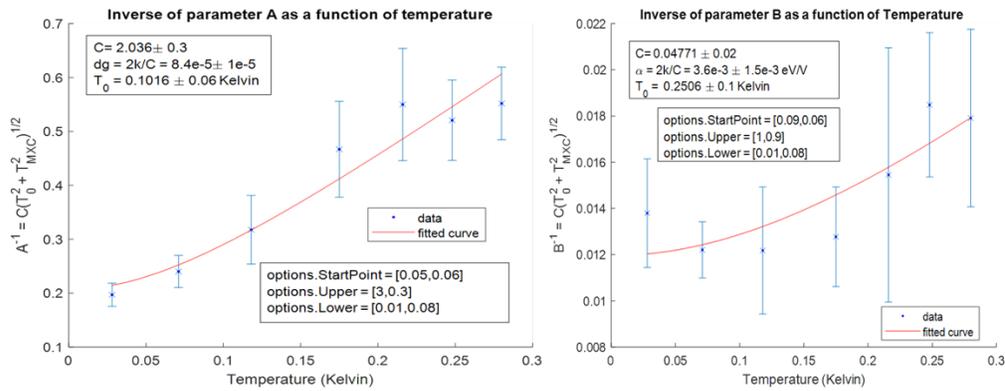

**FIG. S4. Measurement of width of the detuning peaks at different temperatures to get lever arms for the fitting function.**

Using the obtained lever arms, we calculated interdot coupling rate at different $J_g$ voltages.

## IV. PAULI SPIN BLOCKADE IN TRANSPORT AND CHARGE SENSING

We measured p-type transport current and n-type charge sensor current simultaneously at a finite source-drain voltage of 3 mV. We were able to confirm the PSB region from these measurements.

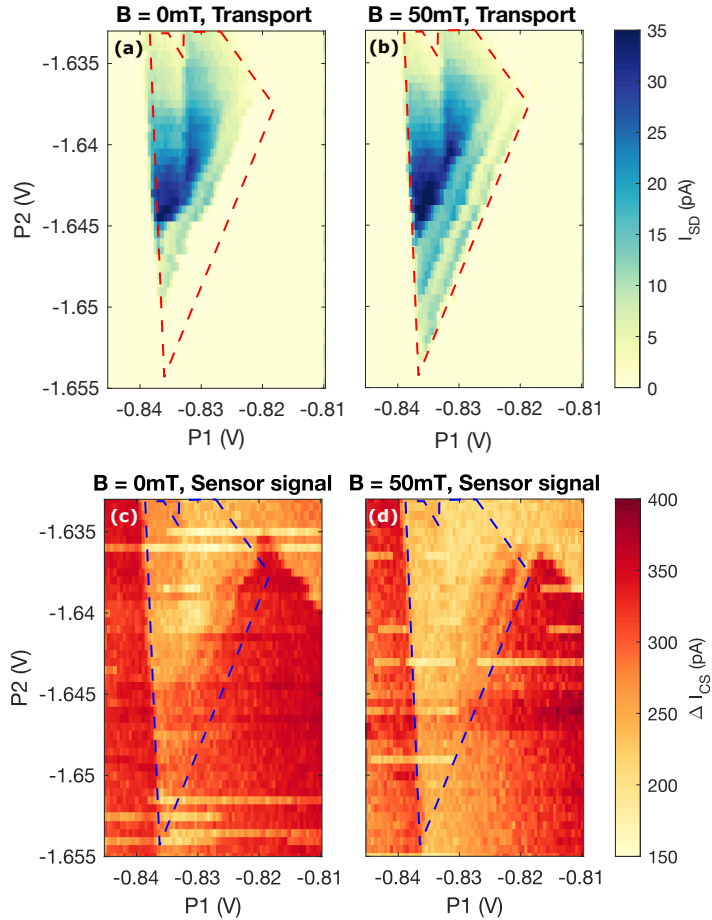

**FIG. S5. Pauli spin blockade measured by p-type transport and n-type charge sensing.** (a) A map of transport current through the p-type double quantum dot shows a bias triangle with suppressed base, (b) whereas a map of the same region with 50 mT of in-plane magnetic field applied shows a clear base peak. (c) The same region is simultaneously mapped with the n-type charge sensor. At zero magnetic field the base of the triangle measures an effective (0,2) charge state (d) whereas at 50 mT the same region measures an effective (1,1) state.

# V. STABILITY DIAGRAM WITH A PULSE SEQUENCE AT DIFFERENT MAGNETIC FIELDS

Stability diagram mapped with a pulse sequence at different magnetic fields. A cyclic three-stage pulse sequence (E-L-M) is applied to measure the transition from the effective (1,1) state to the effective (2,0) state.

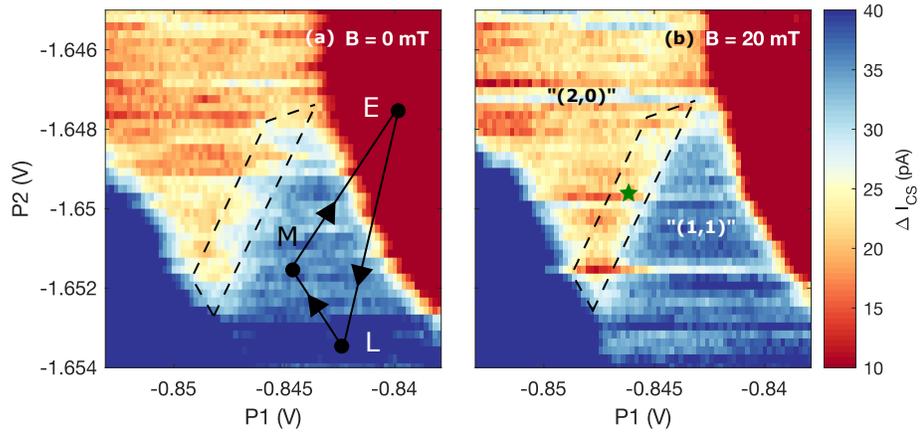

**FIG. S6. Stability diagram with a pulse sequence at different magnetic fields.** (a) When the measurement point is located at the interdot transition region (zero detuning), the triplet "(1,1)" to the singlet "(2,0)" transition is forbidden, which results in PSB (black lines are the guide to the eye). (b) The PSB is lifted when 20 mT of in-plane magnetic field is applied to the double hole quantum dot system.

# VI. SINGLET-TRIPLET RELAXATION TIME CALCULATION

To measure a singlet-triplet relaxation time ($T_{ST}$), we use the forward and reverse pulse and measure the lock-in signal ($I_{CS, AC}$) as a function of pulse period. The data is collected as 144 separate sweeps (Figure S7b is an example of the data measured at zero magnetic field) along the (2,8)-(1,1) detuning axis (white line on Figure S7).

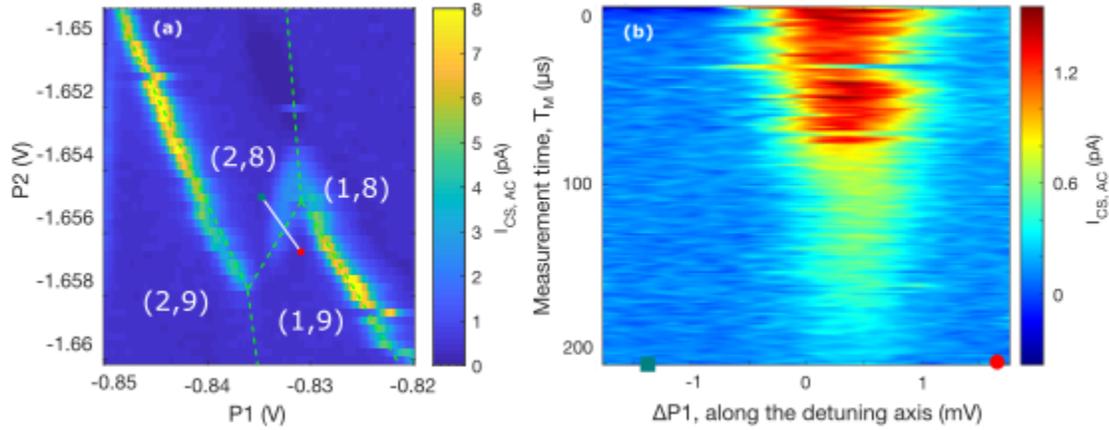

**FIG. S7. Detuning sweeps to obtain singlet-triplet relaxation time.** (a) Stability diagram around (1,9) - (2,8) region. The detuning sweeps are taken along the white line on Figure S7a. (b) Detuning sweeps at zero magnetic field with different pulse periods to extract singlet-triplet relaxation time.

Both two plunger gates P1 and P2 are used to execute a diagonal sweep across the inter-dot region. The bright red in Figure S7b is the spin blocked signal. As the pulse period (y-axis) is made longer, the PSB signal is lost due to singlet-triplet relaxation. We extract the maximum lock-in signal of each detuning sweep, which is plotted in Figure 3d in the main text.

The sensor signal was obtained using integration time of 300 ms, hence the signal results in the average over the entire pulse cycle (*37*). We can model the sensor current by considering a simple exponential decay during the measurement stage, with a characteristic time scale of $T_{ST}$. Hence the extracted maximum current $I_{AVG}$ is given by

$$I_{AVG} = \int_0^{T_E} \frac{I_E}{T} dt + \int_0^{T_R} \frac{I_R}{T} dt + \int_0^{T_M} \frac{I_M}{T} dt$$

$$= A + B \int_0^{T_M} \frac{\exp(-0.75t/T_{ST})}{T} dt$$

$$= A + \frac{BT_{ST}}{T}(1 - e^{-0.75t/T_{ST}})$$

where $I_E$, $I_R$, and $I_M$ are current at the points E, R, and M respectively, $T = T_E + T_R + T_M$ is the total period of the three-stage pulse sequence, and A and B are constants. A factor of 0.75 comes from the fact that the measurement phase (M) takes three quarters of the total period.

We use MATLAB software to fit the data using the command:

myfittype = fittype('a1+b1*(c1/x)*(1-exp(-0.75*x/c1))', 'dependent', ...
'y', 'independent', 'x', 'coefficients', 'a1','b1','c1');

where a1 is the background signal in "(2,0)" (typically around 0.3 to 0.4 pA based on AC background signal level), b1 is the additional signal due to being spin blocked in "(1,1)" (typically around 2.5pA), and c1 is the singlet-triplet relaxation time. The result is summarised in Table I.

|       | 0 mT              | 2 mT              | 5 mT               |
|-------|-------------------|-------------------|--------------------|
| $a_1$ | $0.426 \pm 0.056$ | $0.214 \pm 0.015$ | $0.192 \pm 0.0068$ |
| $b_1$ | $2.01 \pm 0.151$  | $1.77 \pm 0.091$  | $1.32 \pm 0.104$   |
| $c_1$ | $11.1 \pm 2.44$   | $3.81 \pm 0.42$   | $1.22 \pm 0.166$   |

**TABLE S1. Calculated fitting parameters with 95% confidence bounds.** The parameter c1 is equal to the singlet-triplet relaxation time $T_{ST}$.

## VII. SMOOTH CONTROL OF RESERVOIR TUNNELLING RATE

We demonstrate the reservoir tunnelling rate control using a P1 single dot measured by the n-type charge sensor. We applied a periodic pulse sequence at P1 (Figure S8c) to load and unload a hole via the reservoir (Fig. S8a, b).

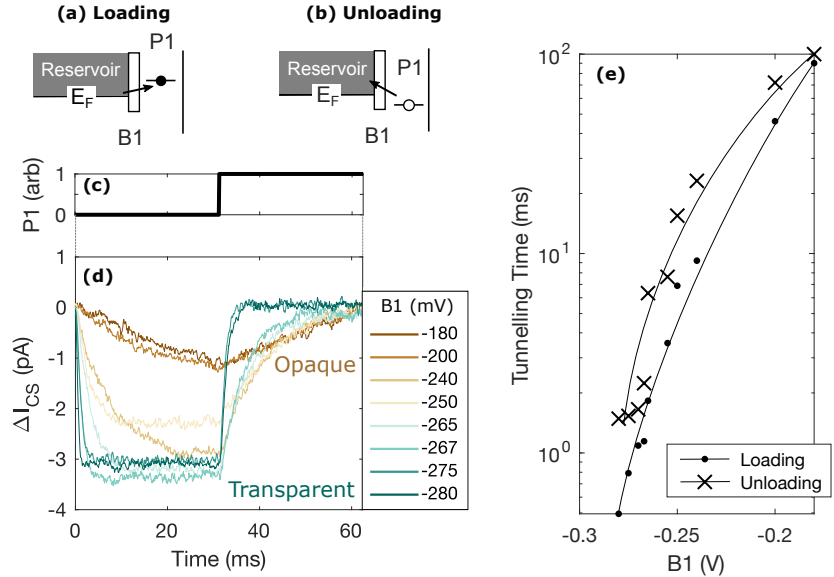

FIG. S8. **Smooth electric control of reservoir tunneling rate.** Changing P1 plunger voltage to (a) load and (b) unload a hole. (c) Pulse sequence applied on P1 plunger gate to measure the reservoir tunnelling rate of the device. (d) Averaged charge sensor current over the periodic pulse sequence. Loading and unloading events of a hole are measured at different reservoir barrier gate voltages. (d) Tunnelling rates of the holes to the reservoir at different reservoir barrier gate voltages, showing the smooth control of the reservoir tunneling rate. Solid lines are guide to the eye.

The n-type charge sensor current ($I_{CS}$) measures the average number of holes we have in the double dot. As a result, the falling (rising) time of $I_{CS}$ corresponds to the loading (unloading) time of a hole. The loading and unloading time are measured at different reservoir gate voltages B1.

Figure S8d depicts time traces when we load and unload a hole at different reservoir barrier voltages. At B1 = −280mV (green), loading time is less than 1 ms; on the other hand, at B1 = −180mV (brown), hole is not fully loaded after the full pulse cycle. Figure S8e summarizes the result, which shows that B1 voltage can smoothly vary the reservoir tunnel coupling rates. The results show that loading time and unloading time can be changed by at least two orders of magnitude.

## VIII. CAPACITANCE MATRIX

Cross-capacitance of the components of the quantum dot device is obtained by electrostatic modelling. An n-type electron dot under the sensor top gate ST and a p-type dot under the plunger gate P2 is denoted by ndot and pdot respectively. The simulation predicts a reasonable capacitive coupling between ndot and pdot, as well as between the dots and the metal gates.

|  | ST | CB1 | SRB | SLB | CB2 | PR1 | PR2 | B1 | P1 | $J_g$ | P2 | B2 | $p_{dot}$ | $n_{dot}$ | Source | Drain |
|---|---|---|---|---|---|---|---|---|---|---|---|---|---|---|---|---|
| $p_{dot}$ | 0.0862 | 1.19 | 0.0662 | 0.0639 | 1.68 | 0.23 | 0.39 | 0.131 | 0.328 | 0.184 | 1.41 | 0.303 | 7.67 | 0.461 | 0.605 | 0.548 |
| $n_{dot}$ | 2.58 | 2.76 | 1.65 | 1.52 | 0.741 | 0.233 | 0.346 | 0.0896 | 0.17 | 0.0532 | 0.162 | 0.117 | 0.461 | 17.5 | 3.36 | 3.27 |

**TABLE S2. Calculated capacitance between the components of the ambipolar CMOS quantum dot device.** For example, the number at (ST, $p_{dot}$) represents the capacitance between the p-type dot and the ST metal gate; ($p_{dot}$, $p_{dot}$) represents self-capacitance of the p-type dot. Units in aF.